\documentclass[12pt]{article}
\usepackage[dvips]{graphicx}
\usepackage{amssymb, amsmath, mathbbol}
\usepackage[centerlast,footnotesize]{caption2}

\usepackage{graphicx}
\newcommand{\One}{1\kern-4.5pt1}

\newcommand{\be}{\begin{equation}}
\newcommand{\ee}{\end{equation}}

\def\lesim{${\lower 2pt\hbox{$\scriptstyle
<$}\atop\raise 4pt\hbox{$\scriptstyle\sim$}}$} 
\def\grsim{${\lower2pt\hbox{$\scriptstyle >$} \atop\raise4pt\hbox 
{$\scriptstyle\sim$}}$} 

\begin{document}
\begin{center}
\begin{flushright}
April 2010
\end{flushright}
\vskip 10mm
{\LARGE
Chiral symmetry breaking and monopole dynamics in non-compact
\medskip
QED$_3$ coupled to a four-fermi interaction
}
\vskip 1.3 cm
{\bf Wesley Armour$^a$, John B. Kogut$^b$ and Costas  Strouthos$^c$}
\vskip 0.3 cm
$^a${\em Diamond Light Source, Harwell Campus,\\
Didcot, Oxfordshire OX11 0DE, U.K.}
\vskip 0.3 cm

$^b${\em Department of Energy, Division of High Energy Physics, \\
Washington,DC 20585, USA \\
and \\
Department of Physics -- TQHN, University of Maryland, 82 Regents Drive, \\
College Park, Maryland 20742, USA.} 

\vskip 0.3 cm
$^c${\em Computation-based Science and Technology Research Center,  \\
The Cyprus Institute,\\
1645 Nicosia , Cyprus.}
\vskip 1.5 cm
\end{center}

\begin{center}
{\bf Abstract} 
\end{center}
\noindent
We present results from the first lattice simulations of three dimensional non-compact 
quantum electrodynamics (QED$_3$) with $N_f$ four-component fermion flavors coupled 
to a weak $Z_2$ chirally invariant four-fermi interaction.
Results with $N_f \ge 4$ show that the 
scaling near the strong coupling chiral transition or sharp crossover  
is determined by the $3d$ Gross-Neveu 
ultraviolet-stable renormalization group fixed point. Small deviations
of the $N_f=4$ critical exponents from the respective Gross-Neveu ones, 
hint at evidence for 
non-zero fermion mass generated by the gauge fields dynamics that might have been enhanced 
by the four-fermi coupling. 
It is also shown that the scaling region is suppressed at weak four-fermi couplings 
and large $N_f$ values.
Measurements of i) a monopole susceptibility which is the polarizability of the monopole 
configurations, and ii) the density of isolated monopoles, 
imply that for $N_f \ge 1$ and weak gauge couplings the monopoles 
do not affect the theory's confining properties, because they are shielded.

                                                                                
\newpage

\section{Introduction}

Spontaneous chiral symmetry breaking plays a significant role in both particle and condensed matter
physics. Of particular interest is the study of quantum field theories in which the ground state 
shows a sensitivity to the number of fermion flavours $N_f$.
Three dimensional parity-invariant quantum electrodynamics with four-component spinors 
is such an interesting and challenging field theory
with rich dynamics that resemble four-dimensional QCD and walking technicolor theories \cite{walking}.  
The gauge coupling $e^2$ has mass dimension one and thus
provides the theory with a natural scale that plays a role similar to $\Lambda_{\rm QCD}$ in four dimensions.
This implies asymptotic freedom,
since for processes with momentum transfer $k\gg e^2$
the theory is effectively non-interacting. 

Non-trivial behavior may arise in
the infrared limit, as suggested by an expansion in $1/N_f$ \cite{pisarski}.
The theory is believed to exhibit logarithmic confinement of electric charges and 
chiral symmetry breaking when the number of fermion
flavors $N_f$ is smaller than a critical value $N_{fc}$. 
Super-renormalizability ensures that QED$_3$ is free from ultraviolet divergences,
thus making analytical calculations
more transparent. Most analytical approaches,
mainly based on self-consistent solutions of
Schwinger-Dyson equations (SDE) 
converge to values of $N_{fc}$ between three and five \cite{Appelquist:1988sr, Goecke:2008zh}. 
There are also SDE results which claim that chiral symmetry is broken for all
values of $N_f$ \cite{pennington}. 
Also, a perturbative analysis of renormalization group
flow in the large-$N_f$ limit
predicts $N_{fc} \approx 6$ \cite{Kaveh:2004qa}. 
Recent progress on gauge invariant solutions of SDE has
been recently reported in \cite{Bashir:2009fv}.
An argument based on the inequality $f_{IR} \leq f_{UV}$
(where $f$ is the absolute value of the thermodynamic free energy) that can be estimated by counting
relevant degrees of freedom in the infrared and ultraviolet limits yields the prediction
$N_{fc}\leq{3\over2}$ \cite{appelquist99}; a result that was later challenged
in \cite{mavromatos}. Analytical calculations
also predict that at $N_{fc}$ the theory undergoes a conformal phase transition \cite{conformal}, which is
a generalization of the infinite order Berezinskii-Kosterlitz-Thouless transition in two dimensions.

Lattice simulations provided evidence  that chiral symmetry is broken for
$N_f < 1.5$ \cite{Hands:1989mv, Hands2004, Strouthos:2008hs},
whereas $N_f=2$ appeared chirally symmetric \cite{Strouthos:2008hs, Hands:2002dv} on lattices
with physical extent up to $Le^2 \approx 90$.
The principal obstruction to a definitive answer has been
the separation of scales in the theory, i.e. the fermion dynamical mass is at least
an order of magnitude smaller than the natural scale $e^2$ \cite{bashir} which is of the
order of the momentum cut-off given by the inverse lattice spacing. In addition,
large finite volume effects resulting from the presence of a massless photon in
the spectrum, prevent a reliable extrapolation to the thermodynamic limit.
Analytical results claimed that to detect chiral symmetry breaking for $N_f \ge 1.5$
lattice volumes much bigger than the ones currently used in numerical simulations
are required \cite{Goecke:2008zh, Gusynin:2003ww}.
Another possibility is that, based on universality arguments,
the infrared limit of QED$_3$ may be equivalent to the $3d$ Thirring model at its
strong coupling ultraviolet-stable renormalization group fixed point, 
as both models are chirally invariant under the same
$U(2N_f)$ group. This universality argument can be valid provided in the Thirring model the
interaction is mediated by a massless
vector boson \cite{Hands:1994kb}, which still needs to be checked rigorously in lattice simulations.
So far, numerical simulations of the Thirring model predicted 
that $N_{fc}=6.6(1)$ \cite{Christofi:2007ye}.

The existence of a critical number of flavors $N_{fc}$ in QED$_3$ can be explained by the following 
semi-classical arguments discussed in \cite{Dagotto:1989td} and also reviewed in \cite{Hands2004}.
In the large-$N_f$ limit
the photon propagator
is modified by the vacuum polarisation diagram
from $1/k^2$ to $1/[k^2+{g^2\over8}N_fk]$ \cite{Appelquist:1986fd}. 
The super-renormalizable theory is rapidly damped in the ultraviolet regime 
and all interesting dynamics are expected in the infrared limit, 
where the dimensionless interaction strength scales as $1/N_f$. 
Naively, one could deduce that the confining property of the Coulomb potential
is screened by virtual fermion-antifermion pairs, because in the
coordinate space the interaction is modified to $1/r$ for distances $r\gg (e^2N_f)^{-1}$.
However, as discussed in  \cite{Dagotto:1989td} the kinetic energy of fermion-antifermion pairs 
is positive and scales as $r^{-1}$ by the uncertainty principle.
In the infrared, therefore, both kinetic and potential terms
scale as $r^{-1}$ and it becomes a delicate question which dominates. 
Since when $r\to0$ the positive kinetic term must
dominate the logarithmic Coulomb term, we deduce in this case the existence of
an energy  minimum at some non-zero $r$,
implying the existence of stable fermion-antifermion bound states in the ground
state. This semi-classical argument suggests a non-vanishing
chiral condensate for  $N_f<N_{fc}$.
For asymptotically large $r$ the massive fermions decouple, ceasing to
screen the charge, and the logarithmically confining Coulomb potential is restored.
For $N_f>N_{fc}$ the theory is conformal, consisting of massless fermions interacting
via a $1/r$ potential. 

Since the early 1990's, non-compact QED$_3$ with or without extra four-fermi terms has 
attracted attention  
\cite{Dorey:1991kp, Gusynin:2000zb} because of
potential applications to models of high $T_c$ superconductivity.
More recently, interest in this model has been revived by suggestions that QED$_3$ 
with two fermion flavors may be an effective theory
for the  underdoped and non-superconducting region of the phase diagram of high-$T_c$ superconducting
cuprate compounds \cite{Franz:2001zz}. In this sense, the abstract theoretical problem of
the value of $N_{fc}$ assumes phenomenological importance.
These results have also stimulated  lattice simulations
of QED$_3$ with Fermi and gap anisotropies \cite{Thomas:2006bj}.

In this paper we present the first exploratory lattice simulation results of 
parity-invariant non-compact QED$_3$ with massless
fermions. This is achieved with the introduction of a four-fermi interaction in 
the QED$_3$ action. 
The lagrangian for the continuum Euclidean field theory
is given by:
\begin{equation}
{\cal L}=\bar\psi_i(\partial{\!\!\!/\,}-i e \gamma_{\mu} A_{\mu} + m)\psi_i-{g_s^2\over{2N_f}}(\bar\psi_i\psi_i)^2+\frac {1}{4} F_{\mu\nu}F_{\mu\nu},
\end{equation}
with the index $i$ implying a summation over $N_f$ four-component fermion flavors. The introduction
of the four-fermi
interaction reduces the $U(2N_f)$ chiral symmetry of QED$_3$ to a discrete $Z_2$ 
symmetry: $\psi_i \to \gamma_5 \psi$; $\bar{\psi_i} \to -\bar{\psi}\gamma_5$.
For computational purposes it is useful to introduce the auxiliary field 
$\sigma \equiv g_s^2\bar{\psi_i}\psi_i$,
and the semi-bosonized lagrangian becomes:
\begin{equation}
\label{eq:bosonized}
{\cal L}=\bar\psi_i(\partial{\!\!\!/\,}-i e \gamma_{\mu} A_{\mu} + \sigma +m)\psi_i
+{N_f \over{2g_s^2}}\sigma^2 +\frac {1}{4} F_{\mu\nu}F_{\mu\nu}.
\end{equation}
Both the non-compact \cite{Kim:2001am} and compact \cite{Kogut:2002vu} 
lattice versions of this model have been used successfully  
to show that QED$_4$ is a logarithmically trivial theory and the systematics of the 
logarithms of triviality follow those of the NJL model rather than those of the scalar $\lambda \phi^4$ theory. 
In the formulation of eq.~(\ref{eq:bosonized}) the $\sigma$ field acts as a chiral 
order parameter which receives a vacuum expectation value, proportional to the chiral condensate, in 
the chirally broken phase. The Dirac operator is now non-singular even with $m=0$ and its inversion is 
very fast. 

The three- and four-dimensional
versions of the theory are very different from each other, as in $3d$ the four-fermi
interaction is a relevant operator. It is well-known that the $3d$ Gross-Neveu model (GNM$_3$), 
although non-renormalizable in weak coupling perturbation theory, 
becomes renormalizable in the $1/N_f$ expansion \cite{GNM}.  
At sufficiently strong couplings and large-$N_f$ 
chiral symmetry is spontaneously broken in GNM$_3$, leading to a dynamically generated fermion mass 
$\Sigma=\langle \sigma\rangle >> m$. The critical coupling $g_{sc}^2$ at which the gap 
$\Sigma/\Lambda_{UV} \to 0$ defines an ultraviolet-stable renormalization group fixed 
point at which an interacting continuum limit may be taken.
As the gauge coupling is varied and the four-fermi coupling is fixed at some value $g_s^2<g^2_{sc}$, then 
depending on the value of $N_f$ the model 
is  expected to undergo either 
a chiral phase transition or a sharp crossover from a 
strong coupling phase (where $\langle \bar{\psi} \psi \rangle \neq 0$) 
to a weak coupling phase where $\langle \bar{\psi} \psi \rangle$
is either zero or very small and possibly undetectable in current lattice simulations.
Hereafter, we will use the term ``chiral transition''
to denote either a chiral phase transition or a sharp crossover from strong to weak gauge couplings.
Unlike what happens in $4d$, here near the transition, the weak four-fermi 
term is expected to  play a dominant role as compared to the ultraviolet-finite
gauge interaction. Understanding the role of the weak four-fermi coupling in 
lattice simulations of non-compact QED$_3$ is one
of the main themes of this paper. Preliminary results were presented in \cite{lat04}. 
Emphasis is also placed  
on the dynamics of monopoles (which in $3d$ are instantons) at both strong and weak 
couplings and try to understand whether they
affect the confining properties of the model.  
The interplay of fermions and magnetic monopoles was 
recently studied in numerical simulations of the compact lattice version of this model
 \cite{Hands:2006dh}.
The authors of \cite{Hands:2006dh} provided evidence that for $N_f=4$ 
the monopole plasma persists even
at weak gauge couplings.
In non-compact QED$_3$, however, the role of topological excitations
may be different,
because in this case  the Dirac strings carry a non-vanishing contribution to
the pure gauge part of the action \cite{Hands:1989cg}. 

The paper is organized as follows. 
In Section~2, we introduce the staggered fermion lattice action and the monopole observables. 
In Section~3 we present results related to the strong coupling chiral transition
such as estimates of the critical exponents $\beta_m$ and $\delta$ for $N_f=4$ and show that they are 
close to the respective GNM$_3$ 
values, implying that the transition is dominated by the ultraviolet-stable GNM$_3$ fixed point. 
Small but systematic deviations of the values of $\beta_m$ and $\delta$ from the GNM$_3$ exponents 
hint at preliminary evidence of nonzero
fermion mass generated by the gauge field dynamics at weak gauge couplings.
We also show that the scaling region is suppressed at weak four-fermi couplings 
and large $N_f$.  
Subsequently, we show that at strong gauge couplings the monopole density has a weak dependence 
on the values of the four-fermi coupling. 
We then study the behavior of the monopole susceptibility $\chi_m$ (defined by Cardy 
in \cite{Cardy:1980jg}) for $N_f=1,...,12$ 
and within the accuracy of our data we observe no diverging behavior in $\chi_m$ with 
the lattice volume, implying that 
monopoles and antimonopoles are shielded in the continuum limit. 
Further evidence in favor of this scenario is provided
by the density of isolated monopoles (positive magnetic charges that don't have 
any anti-monopoles in the nearest neighborhood) which decays faster with 
the inverse gauge coupling than the 
total density of positive magnetic charges. 
In Section~4,  we summarize and discuss our main findings and also point to 
possible future extensions of this work. 

\section{Lattice Model and Observables}
In this first exploratory study of non-compact QED$_3$ with a four-fermi term, we have chosen  
the simplest $Z_2$ chirally invariant four-fermi interaction which for practical purposes 
is preferable over terms with a continuous chiral symmetry, because the latter are not
as efficiently simulated due to massless modes in the strongly cut-off theory. 
The lattice action using staggered lattice fermion fields
$\chi,\bar{\chi}$
is given by the following equations:
\begin{equation}
\label{eq:lattice.action}
S=\frac{\beta}{2}\! \sum_{x,\mu<\nu}\! F_{\mu \nu}(x)\! F^{\mu \nu}(x)\! 
+ \sum_{x,x^\prime} {\bar \chi}(x) Q(x,x^\prime) \chi(x^\prime) 
+\frac{N_f \beta_s}{4} \sum_{\tilde{x}} \sigma^2(\tilde{x}),  
\end{equation} 
where
\begin{eqnarray}
F_{\mu \nu}(x) &\equiv& \alpha_\mu(x)+\alpha_\nu(x+\hat \mu)
-\alpha_\mu(x+\hat \nu)-\alpha_\nu(x), \\
Q(x,x^\prime)&\equiv& 
\frac{1}{2}\! \sum_\mu\eta_{\mu}(x)
[\delta_{x^\prime,x+\hat \mu} U_{x\mu}
\!-\!\delta_{x^\prime,x-\hat \mu} U_{x-\hat \mu,\mu}^\dagger]
+ \delta_{xx^{\prime}}\frac{1}{8}\sum_{\langle \tilde{x}, x \rangle} \sigma(\tilde{x})+m \delta_{xx^{\prime}}.
\end{eqnarray}
The indices $x,~x^\prime$ consist of three integers
$(x_1,~x_2,~x_3)$
labelling the lattice sites, where the third direction is considered
to be timelike.
The symbol $\langle \tilde{x}, x \rangle$ denotes the set of the eight dual lattice sites $\tilde{x}$
surrounding the direct lattice site $x$.
Since the pure gauge action is unbounded from above, eq.~(\ref{eq:lattice.action})
defines the {\sl non-compact\/} formulation of lattice QED.
The $\eta_\mu(x)$ are the Kawamoto-Smit staggered fermion phases $(-1)^{x_1+\cdots+x_{\mu-1}}$,
designed to ensure relativistic covariance of the Dirac equation in the continuum limit.
The boundary conditions for the fermion fields are antiperiodic in
the timelike direction and periodic in the spatial directions.
The phase factors in the fermion bilinear are defined by
$U_{x\mu} \equiv
\exp(i\alpha_{x \mu})$,
where
$\alpha_{x \mu}$ is the gauge
potential.
In terms of continuum quantities,
$\alpha_{x\mu}= a eA_\mu(x)$,
$\beta \equiv \frac{1}{e^2 a}$, $\beta_s \equiv \frac{a}{g_s^2}$
where $a$ is the physical lattice spacing.
Because of the non-compact nature of the gauge fields, the non-fermionic 
part of the action is invariant under gauge transformations defined by the group of real numbers $R$. 
The fermionic
part remains invariant under the smaller gauge group $R/Z \!\sim\! U(1)$. 
The four-fermi term explicitly breaks the $U(N_f/2)\times U(N_f/2)$ chiral symmetry of the lattice
QED$_3$ action to a discrete $Z_2$ chiral symmetry. 

Performing simulations with massless fermions even with the reduced $Z_2$ 
chiral symmetry has substantial advantages, both theoretical and practical.  
The theory has the exact symmetry of the interaction terms, which forbid chiral 
symmetry breaking counter terms from appearing in its effective action. 
In addition, because of the large nonzero vacuum expectation value of the
$\sigma$ field at strong gauge couplings\footnote{At large couplings, pure QED$_3$ simulations are
 dramatically slowed down by the strong gauge field flusctuations.} or 
its fluctuations at weak couplings, 
the model can be simulated very efficiently. 
Another advantage of simulations directly in the chiral limit, is that we do 
not have to rely on often uncontrolled chiral extrapolations to measure the chiral condensate.

The simulations were performed with the standard Hybrid Molecular Dynamics (HMD) R algorithm. 
We used conservatively small values for the HMD trajectory time-step $dt$  
and ensured that any ${\cal O}(dt^2)$ systematic
errors are smaller than the statistical errors on different observables. Among the various parameters
used the ones that are 
the most susceptible to algorithmic systematic errors are those 
near the $N_f=4$ chiral transition ($\beta=0.145$)
on the largest lattice volume $42^3$, 
with the weakest four-fermi coupling $\beta_s=16$. 
By comparing the values of different 
observables obtained from simulations with these parameters 
and with $dt=0.00125$ and $dt=0.0025$ we found agreement within statistical errors.
Therefore, we decided to use $dt=0.0025$ for all $42^3$ and $32^3$ simulations. For the smaller lattices
$16^3$, $24^3$ and $N_f \ge 4$ we found that $dt=0.01$ and $0.005$ are small enough 
to suppress algorithmic systematic
errors at strong coupling and weak gauge couplings, respectively. 
For  $N_f \leq 2$ we used $dt=0.005$ and $dt=0.0025$ for
simulations on  $16^3$ and $24^3$ lattices, respectively.

The magnetic monopoles in the lattice model are identified following the standard  
DeGrand and Toussaint approach \cite{DeGrand:1980eq}.
The plaquette angles $\Theta_{\mu\nu}$ are written as
\begin{equation}
\Theta_{\mu\nu}=\bar{\Theta}_{\mu\nu}+2\pi s_{\mu\nu}(x),
\end{equation}
where $\bar{\Theta}_{\mu\nu}$ lie in the range $(-\pi,\pi]$ and $s_{\mu\nu}(x)$ is an integer that 
determines the flux due to a Dirac string passing through a plaquette. 
The integer number of monopole charges on the dual lattice sites $\tilde{x}$ are then given by 
\begin{equation}
M(\tilde{x})=\epsilon_{\mu\nu\lambda}\Delta_{\mu}s_{\nu \lambda}(\tilde{x}),
\end{equation}
where $\Delta_{\nu}$ is the lattice derivative and $M \in \{0,\pm 1, \pm 2\}$. 
Since on a three-torus the number of monopoles 
is equal to the number of anti-monopoles
we define the density of monopole charges as
\begin{equation}
\rho_M=\frac{1}{V} \sum_{\tilde{x}}|M(\tilde{x})|
\end{equation}
We also measured the monopole susceptibility $\chi_m$ introduced by Cardy \cite{Cardy:1980jg}:
\begin{equation}
\label{eq:susc}
\chi_m=-\frac{1}{V}\sum_{r}\langle r^2 
M(0)M(r) \rangle.
\end{equation}
This observable is the polarizability of the monopole configurations and if the magnetic charges
are in a plasma phase, then  
$\chi_m$ diverges implying that external magnetic fields are shielded. A finite $\chi_m$ implies that  
monopoles and antimonopoles form
tightly bound molecules. The observable $\chi_m$  has been rarely used in simulations with 
dynamical fermions, because it is very noisy due to  
near cancellations of monopole-monopole and monopole-antimonopole contributions. 
With the inclusion of the four-fermi term in the QED$_3$ action the algorithm became very efficient 
and $\chi_m$ has been measured with an acceptable signal-to-noise ratio even at relatively 
strong gauge couplings.

\begin{figure}[t]
    \centering
    \includegraphics[width=10.0cm]{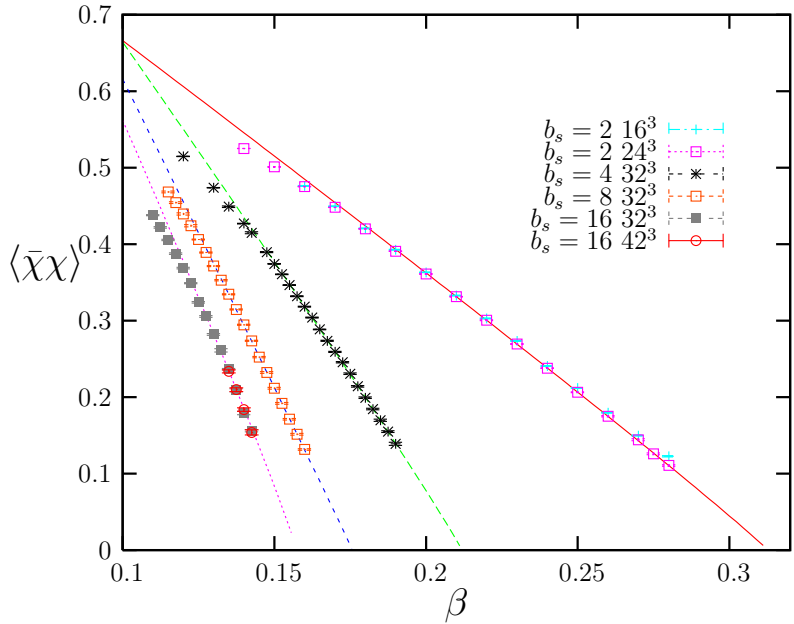}
    \caption{(Color online)
Fits of $\langle \bar{\chi}\chi \rangle$ vs. $\beta$ to eq.~(\ref{eq:scaling}) 
for $N_f=4$ and $\beta_s=2,4,8,16$.}
   \label{gr:fig1}
\end{figure}

\section{Results}
In the infinite gauge coupling limit $\beta \rightarrow 0$, it is known rigorously that chiral symmetry
is broken \cite{salmhofer} for values of $N_f$ below a certain critical value. Simulations of 
QED$_3$ with staggered fermions and $\beta=0$ have shown that the theory undergoes  
a second order phase transition at $N_f \approx 8$ 
with mean field theory exponents \cite{Dagotto:1989td}.
Therefore, as $\beta$ increases, for $N_f > N_{fc}$ there must exist a chiral symmetry
restoring phase transition at some finite $\beta_{c}$.
For $N_f< N_{fc}$, since the order parameter is very small in the continuum limit, 
the relic of the transition may persist as a very sharp crossover
between weak and strong couplings with a tail of an exponentially suppressed
 $\bar\psi\psi$ 
extending to weak gauge couplings\footnote{This may not be directly detectable in lattice simulations.}.
For example, for $N_f=4$, which is the flavor number used in the bulk of our simulations
the SDE approaches predict that the value of the dimensionless chiral        
condensate $\beta^2 \langle \bar{\psi} \psi \rangle$ is somewhere between zero and ${\cal O}(10^{-4})$ 
\cite{Appelquist:1988sr}. In this study we assume that a transition takes place at
some strong (pseudo-)critical gauge coupling $\beta_c$ which depends on $\beta_s$ and we use standard 
scaling relationships for a second order phase transition 
to extract critical exponents. We chose $N_f=4$, instead of a larger $N_f$ value,
because as we show later in this section for large $N_f$ values the width
of the scaling region is suppressed.

\begin{table}[tbp]
\centering \caption{Results from fits of $\langle\bar{\chi}\chi\rangle$ vs. $\beta$ to eq.~(\ref{eq:scaling}).}
\medskip
\label{tab:t1}
\setlength{\tabcolsep}{1.5pc}
\begin{tabular}{ccccc}
\hline \hline
$\beta_s$                & fitting range  &   $\beta_c$    &  $\beta_m$  &  $(\beta_c-\beta_{cross})$   \\
\hline
$2$                      & 0.22-0.28      &  0.313(2)     &  0.96(3)     & 0.09(1)       \\
$4$                      & 0.16-0.19      &  0.212(2)     &  0.96(4)     & 0.052(5)      \\
$8$                      & 0.1325-0.16    &  0.176(1)     &  0.99(3)     & 0.044(3)      \\
$16$                     & 0.1275-0.14    &  0.1579(4)    &  0.93 (fixed) & 0.031(3)     \\
\hline \hline
\end{tabular}
\end{table}

One of the main goals of this first set of lattice simulations of non-compact QED$_3$ 
with the additional four-fermi term
is to understand the impact of the Gross-Neveu coupling on the chiral transition. 
As already mentioned in Section~1, in $3d$ the four-fermi term
becomes a relevant interaction (as opposed to the $4d$ theory where it is an
irrelevant interaction) and therefore this term may play a significant role near the
chiral transition even at very small values of $g_s^2$.
Given that the pure $3d$ Gross-Neveu model with $N_f=4$
undergoes a second order phase transition at $\beta_{sc}=0.835(1)$ \cite{Christofi2006}, 
we chose $\beta_s=2, 4, 8, 16$  for the $N_f=4$ strong gauge coupling simulations.
All these values of $\beta_s$ are in the symmetric phase of the Gross-Neveu model,
implying that they 
cannot generate nonzero fermion dynamical mass on their own.
The finite volume effects are expected to increase with $\beta_s$
and in the limit $\beta_s \to \infty$, on finite size lattices  $\langle \bar{\chi}\chi \rangle \to 0$. 
In order to check the extent of finite volume effects, we performed simulations on $16^3$ and 
$24^3$ lattices for $\beta_s=2$, on $32^3$ lattices for $\beta_s=4, 8$ 
and on $32^3$ and $42^3$ lattices for $\beta_s=16$. 
It is clear from the data for $\langle \bar{\chi}\chi \rangle$ versus $\beta$ 
shown in Fig.~\ref{gr:fig1} that the $24^3$ lattice size
is sufficiently large to suppress finite volume effects for $\beta_s=2$. Also a comparison of the 
values of $\langle \bar{\chi}\chi \rangle$
from simulations with  $\beta_s=16$ on $32^3$ and $42^3$ lattices  implies that 
$32^3$  is large enough to suppress finite size effects for $\beta_s=4, 8, 16$. 
We fitted the data for the different $\beta_s$ values to the standard scaling relation 
of a second order transition order parameter:
\begin{equation}
\label{eq:scaling}
\langle \bar{\chi}\chi\rangle = a(\beta_c-\beta)^{\beta_m}.
\end{equation}
The fitting range was varied to ensure that all the data near the transition that give 
stable values for $\beta_m$ and $\beta_c$ were included. 
For the weakest four-fermi coupling $\beta_s=16$ the three parameter 
fits to eq.~(\ref{eq:scaling}) did
not give stable values for $\beta_m$ and $\beta_c$ when the fitting range was varied.
This is attributed to the fact that for such a very weak four-fermi coupling 
the scaling window is very narrow.
Therefore, in order to obtain an estimate for $\beta_c$ and the width of the
scaling region we fixed $\beta_m=0.93$ (the value of the $3d$ Gross-Neveu exponent \cite{Christofi2006})
and performed two-parameter fits
with eq.~(\ref{eq:scaling}).
A reasonable fit ($\chi^2/{\rm dof}=1.3$) was obtained from the data in the range $\beta=[0.1275-0.140]$.
All the results from this analysis, namely the values for $\beta_c$, $\beta_m$, 
and the fitting ranges (where $\beta_{cross}$
is a crossover coupling that signals the beginning of the scaling region) are 
shown in Table~1. It is deduced that the values of  
$\beta_m$ are in good agreement (or slightly larger by $1-2$ standard deviations) with the 
$\beta_m=0.93(3)$ of the $N_f=4$ GNM$_3$ \cite{Christofi2006}.
An analytical calculation based on SDE and large-$N_f$ approaches for QED$_3$ 
with a $U(1)$ chirally invariant four-fermi term \cite{Mahanta:1991ef} 
predicted that for $N_f > N_{fc}=112/3\pi^2$ the magnetic critical exponent is 
$\beta_m=(3-2a)/4a$, where
$a=\frac{1}{2}\sqrt{1-N_{fc}/N_f)}$. This result reproduces the GNM$_3$  
exponent $\beta_m=1$ only when $N_f \gg N_{fc}$. 

Our results for the critical coupling show that $\beta_c$ decreases as the four-fermi coupling gets weaker
and it is expected to saturate to an asymptotic value 
in the limit $\beta_s \rightarrow \infty$. This asymptotic value is 
clearly smaller than $\beta_c=0.212(4)$
obtained from pure QED$_3$ simulations with $N_f=4$ \cite{Hands2004}. The discrepancy in 
the estimates of $\beta_c$ between the two models 
is attributed to the presence of lattice discretization
counter-terms in the pure QED$_3$ effective action, which are forbidden by the
introduction of the four-fermi term in the current model. 
It should be noted that for the same reason the inclusion of the four-fermi term in the 
non-compact QED$_4$ lagrangian led 
to a larger scaling window in the direction of the 
gauge coupling than in pure non-compact QED$_4$ \cite{Kim:2001am}. 
Another interesting result shown in the last column of Table~\ref{tab:t1} is that the Gross-Neveu scaling 
region $(\beta_c-\beta_{cross})$  is suppressed as the four-fermi coupling 
becomes weaker. This can be understood by a combination of
dimensional analysis and scaling arguments as follows. 
The fixed four-fermi coupling $g_s^2$ with mass dimension $-1$ 
introduces a new scale in the system.
As the scaling region is approached by increasing $\beta$, the  
magnitude of the dimensionless gauge coupling at the crossover into the scaling region 
$e_{\rm cross}^2 a$ ($a$ is the lattice spacing) becomes
comparable to the magnitude of the dimensionless four-fermi coupling $g_s^2 a^{-1}$. Therefore, 
$e_{\rm cross}^2 m_e^2 \sim g_s^2$, where $m_e$ is the electron mass 
which is an inverse correlation length and 
obeys the scaling relation:
\begin{equation}
m_e^{\rm cross} = d(\beta_s)(\beta_c - \beta_{\rm cross})^{\nu},
\end{equation}
with $\nu = 1.0$ for the $N_f=4$ GNM$_3$ \cite{Christofi2006}.
In the large $\beta_s$ limit, $d(\beta_s)$ and $\beta_{\rm cross}$ saturate to 
certain values, implying 
\begin{equation}
\label{eq:width}
(\beta_c-\beta_{\rm cross}) \sim \sqrt{\frac{1}{\beta_s}}.
\end{equation}
It can be seen from the last column of Table~\ref{tab:t1} that the data 
for the scaling region width comply relatively well with eq.~(\ref{eq:width}), 
especially for the two weakest four-fermi couplings $\beta_s=8, 16$.
\begin{figure}[t]
    \centering
    \includegraphics[width=10.0cm]{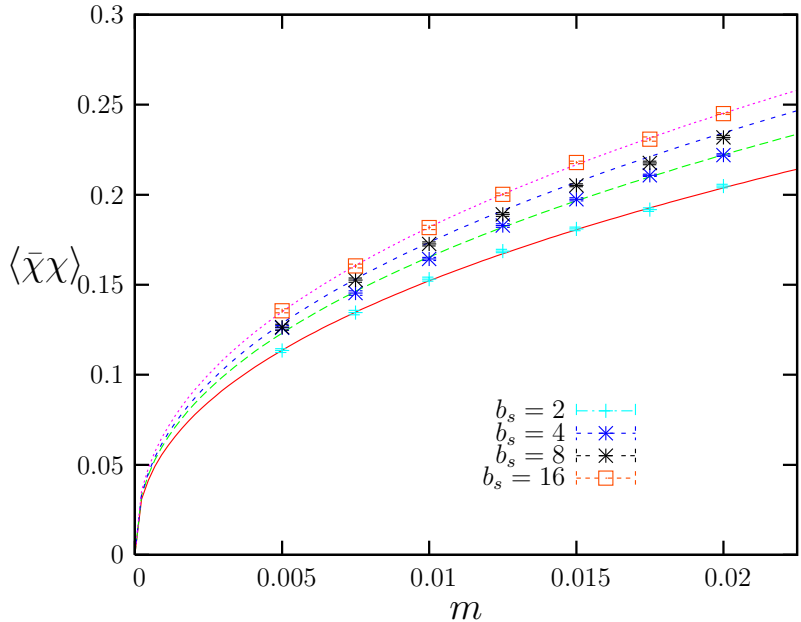}
    \caption{(Color online)
Fits of $\langle\bar{\chi}\chi\rangle$ vs. $m$  to eq.~(\ref{eq:delta}) for $N_f=4$ and $\beta_s=2,4,8,16$.}
   \label{gr:fig2}
\end{figure}

We also performed simulations on $32^3$ lattices with nonzero fermion 
bare mass in the range $m=0.005,...,0.020$ at the respective critical gauge couplings 
for $\beta_s=2, 4, 8, 16$. 
The results for the chiral condensate  as a function of $m$ were fitted to 
the standard scaling relation:
\begin{equation}
\label{eq:delta}
\langle \bar{\chi}\chi\rangle = cm^{1/\delta}.
\end{equation}
The data and the fitting functions are shown in Fig.~\ref{gr:fig2} and the values 
of the critical exponent $\delta$  
are presented 
in Table~\ref{tab:t2}. The extracted values of $\delta$ are close  (within 1-2 standard deviation) 
to the $N_f=4$ GNM$_3$ $\delta=2.24(11)$ \cite{Christofi2006}. The slightly larger values of 
$\delta$ in Table~\ref{tab:t2} as compared to the pure GNM$_3$ value 
could be attributed to either the accuracy with which the
critical couplings were measured and/or to a small fermion mass generated by the gauge field
dynamics. Notably the $\delta=2.45(3)$ extracted from simulations with the strongest  
four-fermi coupling $\beta_s=2$ hints at preliminary evidence in favor of mass generation by 
the QED$_3$ dynamics that might have been enhanced by the four-fermi coupling. 
Future better precision simulations may further clarify this issue. Nevertheless, 
the two critical exponents $\beta_m$ and $\delta$ are sufficient to define 
the universal properties of the model at (pseudo-)criticality, since all the other exponents can 
be estimated using standard hyperscaling relations.
These results clearly indicate that the four-fermi interaction plays a predominant role in the 
scaling of the order parameter near the chiral transition. Although 
the transition may be approached by varying 
the gauge coupling alone, as mentioned in Section~1 pure QED$_3$ 
is super-renormalizable and ultraviolet-finite, and therefore the ultraviolet-stable fixed point of 
GNM$_3$ determines the scaling properties of the transition.
This situation is similar to the 
Higgs-Yukawa model where the interaction $\lambda \phi^4$ is also super-renormalizable in $3d$, 
and the four-fermi interaction determines the model's universality class \cite{Focht:1995ie}.  
An analogous  result was recently obtained in a perturbative 
renormalization group theory approach of the graphene  
phase diagram \cite{Herbut}, where it was shown that in the presence 
of a four-fermi interaction the electron 
charge is rendered more marginally irrelevant than at the Gaussian fixed point.  

\begin{table}[t]
\centering \caption{Results for exponent $\delta$ extracted from fits to eq.~(\ref{eq:delta})
for $N_f=4$.}
\medskip
\label{tab:t2}
\setlength{\tabcolsep}{1.0pc}
\begin{tabular}{cc}
\hline \hline
$\beta_s$               & $\delta$           \\
\hline
$2$                      & $2.45(3)$       \\
$4$                      & $2.36(3)$       \\
$8$                      & $2.34(3)$     \\
$16$                     & $2.33(2)$     \\
\hline \hline
\end{tabular}
\end{table}

\begin{figure}[t]
    \centering
    \includegraphics[width=10.0cm]{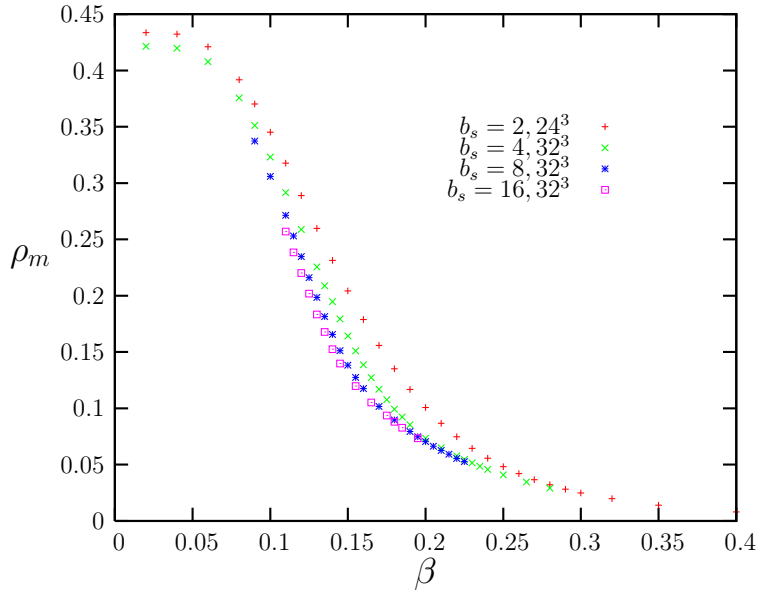}
    \caption{(Color online)
$\rho_m$ vs. $\beta$ for $N_f=4$ and $\beta_s=2,4,8,16$.}
   \label{gr:fig4}
\end{figure}

In Fig.~\ref{gr:fig4} we present the results for the monopole density $\rho_m$ versus $\beta$ 
for $\beta_s=2, 4, 8, 16$. We see that $\rho_m$ decreases 
monotonically and smoothly with $\beta$ with no evidence for any abrupt changes at the chiral 
transition couplings that might signal 
the existence of a phase transition from strong to weak couplings. At strong 
gauge couplings $\rho_m$ decreases as $\beta_s$ increases and above the chiral transition $\rho_m$ tends to 
become $\beta_s$-independent. It should be noted, however, that the impact of $\beta_s$ 
on $\rho_m$ is significantly smaller than 
on the chiral condensate (see Fig.~\ref{gr:fig1}); a similar result was obtained in simulations of 
the compact version of the model \cite{Hands:2006dh}. As expected, a stronger four-fermi coupling 
produces a larger enhancement of the  fermion condensate near the transition, 
whereas the monopole dynamics are mainly determined 
by the short distance fluctuations of the gauge fields. 
\begin{figure}[thbp]
    \centering
    \includegraphics[width=10.0cm]{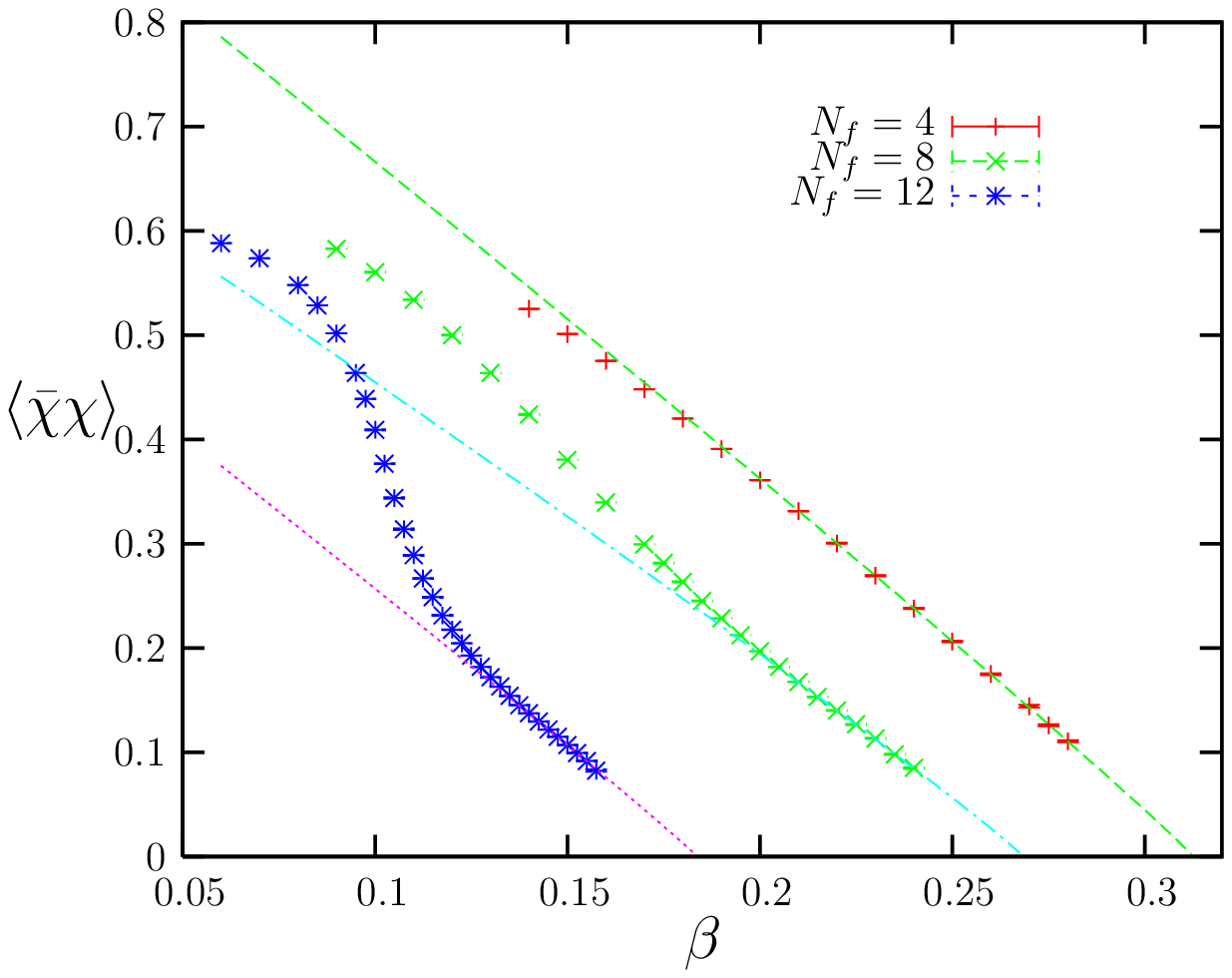}
    \caption{(Color online)
$\langle\bar{\chi}\chi\rangle$ vs. $\beta$ for
$N_f=4, 8, 12$.}
   \label{gr:fig3}
\vspace{8mm}
    \centering
    \includegraphics[width=10.0cm]{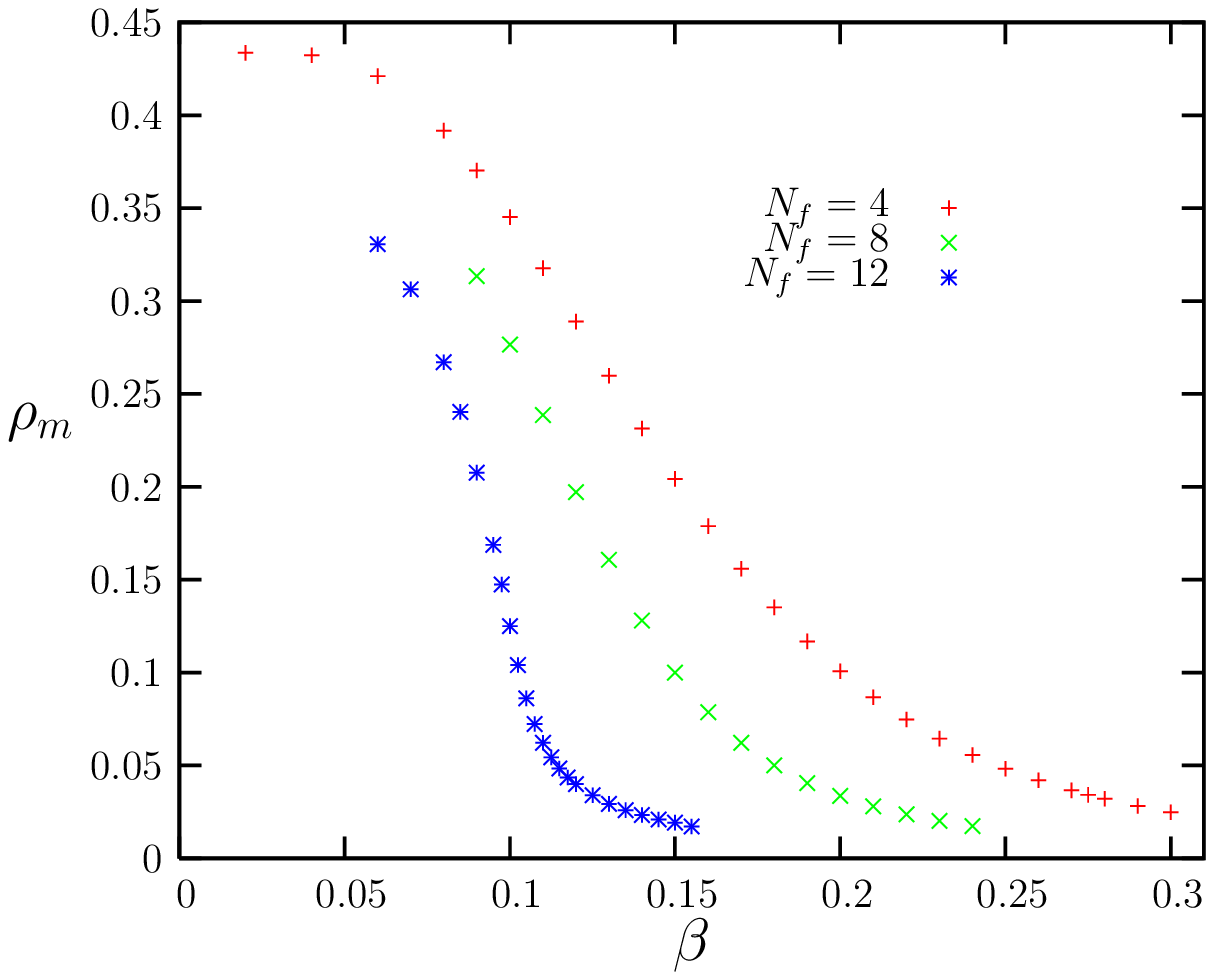}
    \caption{(Color online)
Monopole density $\rho_m$ vs. $\beta$ for $N_f=4, 8, 12$, $\beta_s=2$.}
   \label{gr:fig6}
\end{figure}
\begin{figure}[htbp]
    \centering
    \includegraphics[width=10.0cm]{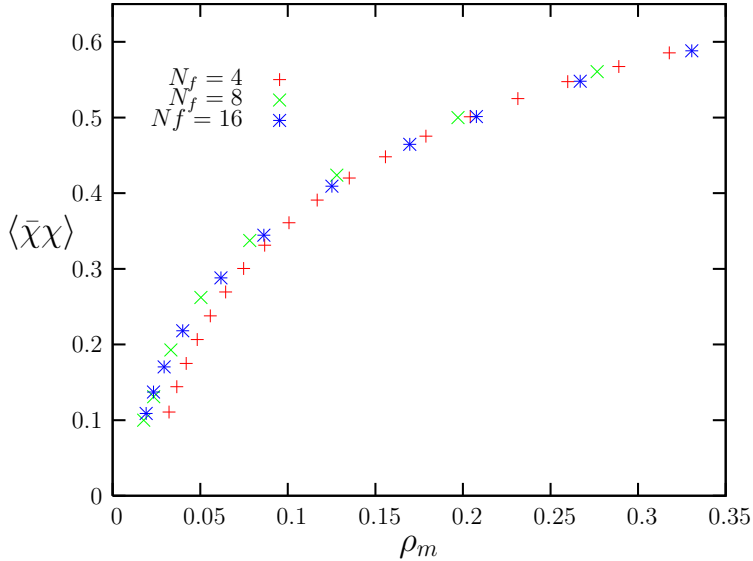}
    \caption{(Color online)
$\langle\bar{\chi}\chi\rangle$ vs. $\rho_m$ for $N_f=4, 8, 12$ and $\beta_s=2$.}
   \label{gr:fig7}
\end{figure}

\begin{table}[tbp]
\centering \caption{Results from fits of $\langle\bar{\chi}\chi\rangle$ vs. $\beta$ 
to eq.~(\ref{eq:scaling}) for $N_f=4,8,12$ and compliance of the width of the scaling region to eq.~(\ref{eq:scaling.nf}).}
\medskip
\label{tab:t3}
\setlength{\tabcolsep}{1.0pc}
\begin{tabular}{cccccc}
\hline \hline
$N_f$    & fitting range  &   $\beta_c$    &  $\beta_m$  &  $R \equiv (\beta_c-\beta_{\rm cross})$ &  $R/\sqrt{\frac{\beta_{\rm cross}(N_f)}{N_f}}$\\
\hline
$4$                      & 0.22-0.28      &  0.313(2)     &  0.96(3)     & 0.093(8) &   0.40(9)   \\
$8$                      & 0.21-0.24      &  0.269(3)     &   0.95(6)    & 0.059(5) &   0.36(7)  \\
$12$                      & 0.140-0.1575   &  0.184(3)     &  0.97(11)     & 0.044(5) &  0.41(6)    \\
\hline \hline
\end{tabular}
\end{table}

In order to study the dependence of the scaling region width on $N_f$ we performed additional 
simulations with  $N_f=8, 12$ and $m=0$, 
$\beta_s=2$ on $24^3$ lattices.
It is shown in Fig.~\ref{gr:fig3} that as $N_f$ increases the values 
of $\langle \bar{\chi}\chi\rangle$ decrease,
due to the enhanced screening of the electromagnetic interaction by virtual fermion-antifermion pairs.
The data for $\langle\bar{\chi}\chi\rangle$ were fitted to eq.~(\ref{eq:scaling}) 
in the regions where stable values for $\beta_m$ and $\beta_c$ were obtained. 
The results for $N_f=8, 12$ together with the $N_f=4$ results discussed earlier, 
and the fitted curves are shown in  
 Fig.~\ref{gr:fig3} and summarized in Table \ref{tab:t3}. 
The values of $\beta_c$ decrease as $N_f$ increases and the values of the magnetic critical exponent 
$\beta_m$ are close to the large-$N_f$ GNM$_3$ value $\beta_m=1$.  
As already mentioned in Section~1 the dimensionless strength of the QED$_3$ 
interaction  
is given by $\sim 1/N_f$.  
Following the reasoning that led to the derivation of eq.~(\ref{eq:width}), 
it can be easily shown that for fixed $\beta_s$ and variable $N_f$ 
the width of the Gross-Neveu scaling
region is given by:
\begin{equation}
(\beta_c-\beta_{\rm cross}) \sim \sqrt{\frac{\beta_{\rm cross}(N_f)}{N_f}},
\label{eq:scaling.nf}
\end{equation}
where $\beta_{\rm cross}(N_f)$ is the crossover coupling into the GNM$_3$ scaling region. 
The results presented in  
Table~\ref{tab:t3} show that the scaling region is suppressed for large $N_f$ values. It is also  
inferred from the results in the last column of Table~\ref{tab:t3} that the data  comply with 
eq.~(\ref{eq:scaling.nf}) relatively well.
In a similar way to the chiral condensate, the monopole 
excitations density shown in Fig.~\ref{gr:fig6} also gets smaller as the interaction strength 
is decreased by increasing $N_f$. 
It is also interesting to observe that in Fig.~\ref{gr:fig7} the data for
$\langle\bar{\chi}\chi\rangle$ versus $\rho_m$ for $N_f=4, 8, 12$ and $\beta_s=2$ collapse on a single 
curve, the only exception being the $N_f=4$ data at small values of $\rho_m$, 
a regime that coincides with the broader $N_f=4$ GNM$_3$ scaling region. 
We infer from Fig.~\ref{gr:fig7} that large $N_f$ values 
and strong gauge couplings produce chiral condensation 
and topological excitations in a very similar almost $N_f$-independent manner, without this implying 
anything about the impact of monopoles on the theory's confining properties. From the monopole density 
alone one cannot reach conclusions regarding the confining properties of the theory.
\begin{figure}
    \centering
    \includegraphics[width=10.0cm]{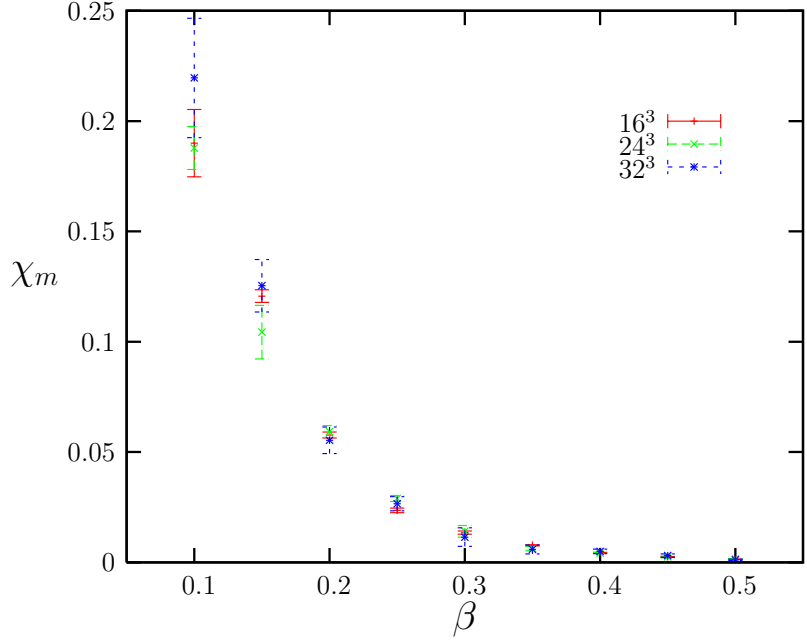}
    \caption{(Color online)
$\chi_m$ vs. $\beta$ for $N_f=4$, $\beta_s=2$ 
on $16^3$, $24^3$, and $32^3$ lattices.}   
   \label{gr:fig5}
\end{figure}
\begin{figure}[thbp]
    \centering
    \includegraphics[width=10.0cm]{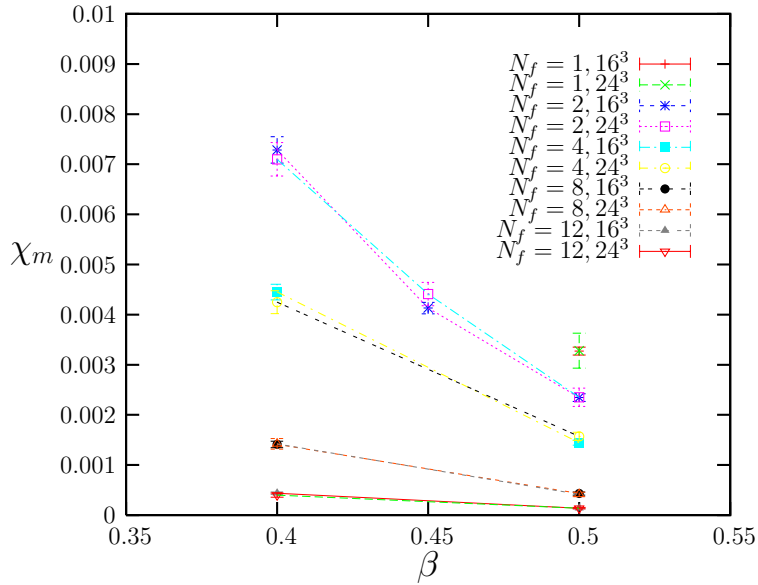}
    \caption{(Color online)
$\chi_m$ vs. $\beta$ for $N_f=1, 2, ...,12$,
$\beta_s=2$ on $16^3$ and $24^3$ lattices.}
   \label{gr:fig8}
\end{figure}
\begin{figure}[htbp]
    \centering
    \includegraphics[width=10.0cm]{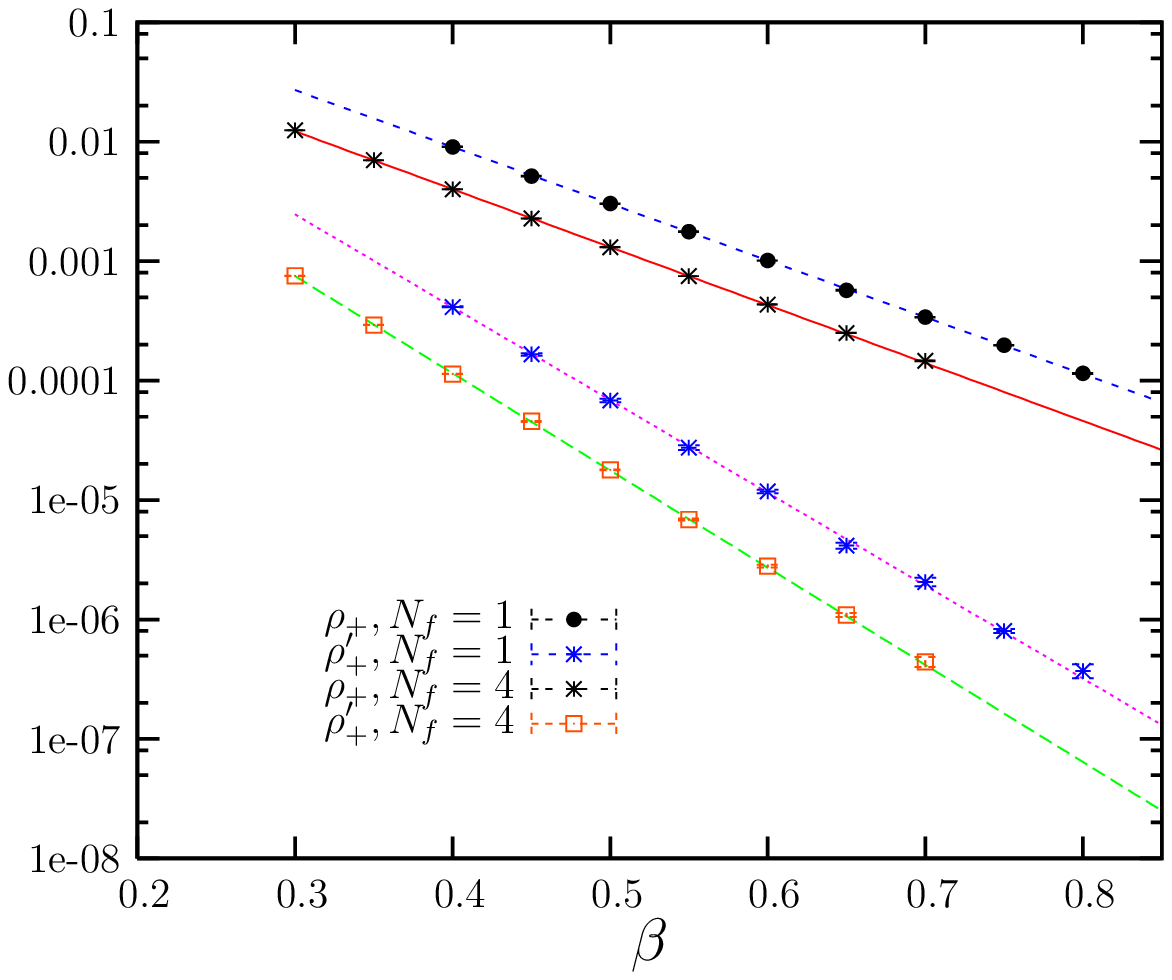}
    \caption{(Color online)
Positive monopole charges density $\rho_+$ and isolated positive monopole charges density $\rho_+^{\prime}$ 
vs. $\beta$ for $N_f=1, 4$, $\beta_s=2$ on $32^3$ lattice.}
   \label{gr:fig9} 
\end{figure}

Next, we take a closer look at the monopole dynamics by discussing results for the monopole 
susceptibility $\chi_m$ defined in eq.~(\ref{eq:susc}). This observable is far more informative than 
$\rho_m$, because it measures the polarizability
of the monopole configurations and it is expected to show a diverging behavior 
if the monopoles and antimonopoles are in a plasma phase. 
As mentioned in Section~2, $\chi_m$ is in general very noisy due to near 
cancellations of monopole-monopole and monopole-antimonopole contributions. 
However, the four-fermi term introduced in the QED$_3$ action 
substantially increased the efficiency of the simulation  algorithm 
and enabled us to measure $\chi_m$ with acceptable signal-to-noise ratio even at relatively 
strong couplings. 
The data for $\chi_m$ versus $\beta$ obtained from simulations with $N_f=4$, $\beta_s=2$ 
on $16^3$, $24^3$ and $32^3$ lattices are presented in Fig.~\ref{gr:fig5}. 
Although $\chi_m$ increases monotonically with the gauge coupling, 
there are no signs of a divergent behavior. 
These results imply that monopoles
are shielded by forming tightly bound molecules. 
We also performed simulations with different numbers of
fermion flavors in the range $N_f=1,2...,12$ on $16^3$ and $24^3$ lattices
with $\beta_s=2$. The values of $\beta=0.4, 0.5$ used are in the weak coupling
phase for all $N_f$ values in the sense that $\langle \bar{\psi}\psi \rangle$ 
is consistent with zero and the effective order
parameter $\langle |\bar{\psi}\psi| \rangle$ decreases with the lattice volume.
Again, the data for $\chi_m$ presented in Fig.~\ref{gr:fig8}
show no signs of a diverging behavior as they are independent
of the lattice volume within statistical errors. These results imply that for $N_f \ge 1$ 
the monopole dynamics do not
affect the confining properties of the theory's continuum limit.

The conclusion of the previous paragraph is strengthened by 
the data in Fig.~\ref{gr:fig9}, where we show for $N_f=1$ and 4
the density of positive magnetic charges (monopoles) $\rho_+$
and the density of monopoles without any antimonopoles in 
their nearest neighborhood ($x_i\pm 1$) $\rho_+^{\prime}$ versus beta. 
We fitted these data to an empirical 
function $f(\beta)=a_1 \exp(-a_2 \beta)$ and the extracted  values 
of $a_2$ and $a_2^{\prime}$ for $\rho_+$ and $\rho_+^{\prime}$, respectively, are shown 
in Table~\ref{tab:t4}.
The fact that $\rho_+^{\prime}$ decays faster than $\rho_+$ confirms the scenario
that at weak couplings the monopoles are shielded.
The increase of $\chi_m$ and $\rho_+^{\prime}$ at strong gauge couplings and/or small $N_f$ values
can be easily understood in terms of the renormalization group invariant Dirac quantization condition
$eg=e_Rg_{R}$ \cite{Calucci:1982wy} ($g$ and $g_R$ are the monopole bare and renormalized charges).
As  $N_f$ (or $\beta$) decreases the 
interaction between fermions and antifermions gets stronger, implying that the 
monopole-antimonopole attraction gets weaker. 

\begin{table}[tbp]
\centering \caption{Results from fits of $\rho_+$ and $\rho_+^{\prime}$ vs. $\beta$ 
to $f(\beta)=a_1 \exp(-a_2 \beta)$  for $N_f=1,4$.}
\medskip
\label{tab:t4}
\setlength{\tabcolsep}{1.5pc}
\begin{tabular}{ccc}
\hline \hline
$N_f$                & 1           &   4           \\
\hline
$a_2$               & 10.9(1)      &  11.16(4)   \\
$a_2^{\prime}$      & 17.90(9)     &  18.74(3)   \\
\hline \hline
\end{tabular}
\end{table}

The monopole dynamics in the compact and non-compact formulations  
may be different at weak couplings and small $N_f$ values, because in the non-compact version 
the Dirac strings carry a non-vanishing contribution to the pure gauge part of the action \cite{Hands:1989cg}. 
The two formulations, however, become identical
in the infinite gauge coupling ($\beta \to 0$) and/or large-$N_f$ ($N_f \to \infty$) limits, as
in these limits the fermionic sector with compact gauge links in the action
dominates the dynamics. The results from simulations of compact QED$_3$ coupled 
to a four-fermi term with $N_f=4$  \cite{Hands:2006dh} provided evidence that 
the monopole plasma phase persists even at weak gauge couplings. Although our simulations  
with $N_f=4$ were performed at relatively strong couplings deep in the chirally broken phase, 
it is unclear whether we reached the limit where the monopole dynamics in the 
two formulations should become similar.
Furthermore, it has been suggested \cite{Fiebig, Fiore:2005ps} that the compact and non-compact formulations 
may be equivalent. This suggestion 
was largely based on 
comparisons of chirally extrapolated data 
for $\langle\bar{\chi}\chi\rangle$ versus $\rho_m$ from the two models. 
The authors of \cite{Fiore:2005ps} also suggested that even at weak couplings the monopoles 
in non-compact QED$_3$
may exist in a plasma phase. 
Our results and their difference from those reported in \cite{Hands:2006dh} 
imply that these suggestions may not be valid.
Marston \cite{marston} and later on Kleinert and collaborators \cite{Kleinert, Nogueira:2007pn} 
suggested in a series 
of analytical papers on compact QED$_3$, 
that above a certain critical number of fermion flavors ($N_{fc}$) a phase exists where 
the monopole-antimonopole potential is modified from $1/r$ to $\ln(r)$, 
implying that monopoles are shielded and do not influence the continuum limit. 
Recently, it was reported that $N_{fc}=36$ \cite{Nogueira:2007pn}.
Also Herbut and collaborators \cite{Herbut:2003bs} 
claimed that in compact QED$_3$, the interaction 
among magnetic dipole pairs restores the Coulomb potential at large distances, and 
the monopoles are in a plasma phase at all $\beta < \infty$ and $N_f < \infty$. 
In the near future we plan to simulate compact QED$_3$ with different $N_f$ values 
and compare the results for monopole dynamics with those mentioned above. 

\section{Summary and Outlook}
\label{sec:discussion}
Three dimensional QED is an interesting field theory due to its similarities to four-dimensional 
QCD-like theories and its applications in high temperature superconductivity. 
In this paper we presented the first results from 
lattice simulations with massless fermions of non-compact QED$_3$ coupled to a weak $Z_2$ 
chirally invariant four-fermi interaction.
Below we summarize and discuss our main findings. 

The values of         
the critical exponents $\beta_m$ and $\delta$ extracted from simulations with $N_f \ge 4$ are close to 
the values of the respective $3d$ Gross-Neveu exponents. This implies that 
even for very weak four-fermi couplings 
(almost twenty times smaller than the pure Gross-Neveu 
critical coupling) the strong coupling chiral transition/crossover 
for $N_f \ge 4$ is dominated by the Gross-Neveu ultraviolet-stable renormalization group fixed point. 
The gauge interaction is an irrelevant operator near the transition, but the system 
can be driven towards (pseudo-)criticality by varying the gauge coupling. 
It is noted that certain values of the critical exponents (especially $\delta$)
extracted from simulations with the strongest four-fermi coupling, $\beta_s=2$, 
are slightly larger than those of the pure Gross-Neveu exponents, 
which hints at preliminary evidence for nonzero fermion mass  
generated by the gauge field dynamics. Future large scale simulations 
will further clarify this issue. 
In the weak four-fermi coupling limit the
$N_f=4$ critical gauge coupling $\beta_c$ is significantly smaller 
than its respective pure $N_f=4$ QED$_3$ value, 
possibly because the four-fermi term forbids lattice discretization counter-terms 
that may exist in the pure QED$_3$ effective action with $m \neq 0$. Furthermore, for a given $N_f$ value, 
in the weak four-fermi coupling limit the scaling  
region is suppressed by a factor $\sim 1/\sqrt{\beta_s}$, whereas for fixed $\beta_s$ 
and variable $N_f$ the scaling region is suppressed by a factor $\sim \sqrt{\beta_{\rm cross}(N_f)/N_f}$. 
Also, the effects of the four-fermi coupling on the monopole density 
are smaller than on the chiral condensate, as $\rho_m$ depends 
more strongly on the short distance fluctuations of the gauge field, whereas the 
fermion condensate is significantly enhanced near the transition 
by the  four-fermi coupling. For $N_f \ge 4$
we have also seen evidence that at strong couplings the chiral condensate is correlated to
the monopole density in an $N_f$-independent manner. 

We also reached the conclusion that the monopoles are shielded at weak couplings 
for $N_f \ge 1$ because: i) the monopole 
polarizability measured by $\chi_m$ showed no signs of a divergent 
behavior, and ii) the exponential decrease of the isolated  monopoles' density with $\beta$ 
is faster than the decrease of the total density of positive magnetic charges. 
A comparison of our results with those from simulations of 
compact QED$_3$ coupled to a four-fermi term \cite{Hands:2006dh} which 
favor survival of the monopole plasma at weak couplings for $N_f=4$
implies that the monopole dynamics in the two models are different.

The various results from this first exploratory study are promising. This is due to   
the presence of the four-fermi coupling in the model's action, which increased 
substantially the efficiency of the simulation 
algorithm. In the near future we plan to extend this work in the following directions: 
i) Large scale simulations of the non-compact version with $N_f \leq 4$ to compare the
scaling properties at the transition/crossover with those of pure Gross-Neveu model. 
Possible deviations
may provide evidence for the existence of fermion mass generated by the gauge field dynamics. 
ii) Landau gauge-fixed simulations to measure the fermion dynamical mass, which according to 
SDE approaches it is expected to be an order of magnitude 
smaller than the natural cut-off scale $e^2$ in the continuum limit \cite{bashir}, 
and hence significantly larger than the chiral condensate. 
iii) Simulations of compact QED$_3$ with different values of $N_f$ and comparison with 
the existing analytical results.

\section*{Acknowledgements}
The authors are grateful to Simon Hands and Pavlos Vranas for valuable insight and discussions.
JBK thanks the National Science Foundation for providing computer time at several NSF supported
computer centers under grant MCA99S015. The authors also wish to thank Diamond Light Source for 
kindly allowing them to use extensive computing resources, specifically Tina Friedrich,
Frederik Ferner, James Rowland and Alun Ashton for help in configuring and
maintaining these resources. WA would like to thank Gwyndaf Evans for his support and useful advice.

\end{document}